\def\papertitle{Pitch-Conditioned Instrument Sound Synthesis\\from an Interactive Timbre Latent Space}
\def\paperauthorA{Christian Limberg}
\def\paperauthorB{Fares Schulz}
\def\paperauthorC{Zhe Zhang}
\def\paperauthorD{Stefan Weinzierl}

\documentclass[twoside,a4paper]{article}
\usepackage{etoolbox}
\usepackage{dafx_25}
\usepackage{amsmath,amssymb,amsfonts,amsthm}
\usepackage{euscript}
\usepackage[T1]{fontenc}
\usepackage[utf8]{inputenc}
\usepackage{ifpdf}
\usepackage[english]{babel}
\usepackage{caption}
\usepackage{subfig} 
\usepackage{color}
\usepackage{flushend}

\input glyphtounicode
\ninept

\newcounter{numauth}
\newcounter{listcnt}
\newcommand\authcnt[1]{\ifdefined#1 \stepcounter{numauth} \fi}

\newcommand\addauth[1]{
\ifdefined#1 
\stepcounter{listcnt}
\ifnum \value{listcnt}<\value{numauth}
\appto\authorslist{, #1}
\else
\appto\authorslist{~and~#1}
\fi
\fi}
\authcnt{\paperauthorB}
\authcnt{\paperauthorC}
\authcnt{\paperauthorD}
\authcnt{\paperauthorE}
\authcnt{\paperauthorF}
\authcnt{\paperauthorG}
\authcnt{\paperauthorH}
\authcnt{\paperauthorI}
\authcnt{\paperauthorJ}
\def\authorslist{\paperauthorA}
\addauth{\paperauthorB}
\addauth{\paperauthorC}
\addauth{\paperauthorD}
\addauth{\paperauthorE}
\addauth{\paperauthorF}
\addauth{\paperauthorG}
\addauth{\paperauthorH}
\addauth{\paperauthorI}
\addauth{\paperauthorJ}

\usepackage{times}

\newif\ifpdf
\ifx\pdfoutput\relax
\else
   \ifcase\pdfoutput
      \pdffalse
   \else
      \pdftrue
   \fi
\fi

\ifpdf 
  \usepackage[pdftex,
    pdftitle={\papertitle},
    pdfauthor={\authorslist},
    pdfsubject={Proceedings of the 28th International Conference on Digital Audio Effects (DAFx25)},
    colorlinks=false, 
    bookmarksnumbered, 
    pdfstartview=XYZ 
  ]{hyperref}
  \usepackage[pdftex]{graphicx}
  \usepackage{orcidlink}
\else 
  \usepackage[dvips]{epsfig,graphicx}
  \usepackage[dvips,
    pdftitle={\papertitle},
    pdfauthor={\authorslist},
    pdfsubject={Proceedings of the 28th International Conference on Digital Audio Effects (DAFx25)},
    colorlinks=false, 
    bookmarksnumbered, 
    pdfstartview=XYZ 
  ]{hyperref}
\fi
\usepackage[hypcap=true]{caption}
\title{\papertitle}

 \fouraffiliations{
 \paperauthorA\ * \,\thanks{* Both authors have contributed equally to this work.}}
 {\href{https://www.tu.berlin/en/ak}{Audio Communication Group} \\ Technische Universität Berlin\\ Berlin, DE\\
 {\tt \href{mailto:dafx25@dii.univpm.it}{cnlimberg@gmail.com}}
 }
 {\paperauthorB\ \orcidlink{0009-0003-3512-0096} * \,\thanks{\vspace{-3mm}}}
 {\href{https://www.tu.berlin/en/ak}{Audio Communication Group} \\ Technische Universität Berlin\\ Berlin, DE\\
 {\tt \href{mailto:contact@faresschulz.com}{fares.schulz@tu-berlin.de}}
 }
 {\paperauthorC \,\thanks{\vspace{-3mm}}}
 {\href{https://yamagishilab.jp/}{Yamagishi Lab} \\ National Institute of Informatics \\ Tokyo, JPN \\
 {\tt \href{mailto:dafx2023@gmail.com}{zhe@nii.ac.jp}}
 }
 {\paperauthorD \,\thanks{\vspace{-3mm}}}
 {\href{https://www.tu.berlin/en/ak}{Audio Communication Group} \\ Technische Universität Berlin\\ Berlin, DE\\
 {\tt \href{mailto:stefan.weinzierl@tu-berlin.de}{stefan.weinzierl@tu-berlin.de}}
 }
\begin{document}
\ifpdf 
  \DeclareGraphicsExtensions{.png,.jpg,.pdf}
\else  
  \DeclareGraphicsExtensions{.eps}
\fi


\maketitle

\begin{abstract}
This paper presents a novel approach to neural instrument sound synthesis using a two-stage semi-supervised learning framework capable of generating pitch-accurate, high-quality music samples from an expressive timbre latent space.
Existing approaches that achieve sufficient quality for music production often rely on high-dimensional latent representations that are difficult to navigate and provide unintuitive user experiences.
We address this limitation through a two-stage training paradigm: first, we train a pitch-timbre disentangled 2D representation of audio samples using a Variational Autoencoder; second, we use this representation as conditioning input for a Transformer-based generative model. The learned 2D latent space serves as an intuitive interface for navigating and exploring the sound landscape.
We demonstrate that the proposed method effectively learns a disentangled timbre space, enabling expressive and controllable audio generation with reliable pitch conditioning. Experimental results show the model’s ability to capture subtle variations in timbre while maintaining a high degree of pitch accuracy.
The usability of our method is demonstrated in an interactive web application, highlighting its potential as a step towards future music production environments that are both intuitive and creatively empowering:\newline
\href{https://pgesam.faresschulz.com/}{https://pgesam.faresschulz.com/}.

\end{abstract}

\section{Introduction}
\label{sec:intro}

The creation and exploration of musical samples utilizing deep learning technologies offer transformative possibilities for music production. Recent advancements in generative audio synthesis prominently feature techniques such as Generative Adversarial Networks (GANs) and Variational Autoencoders (VAEs), which have shown great promise in generating waveform audio and complex musical textures \cite{briot_music_2020, ji_comprehensive_2020-3}. Despite this, these approaches often require extensive fine-tuning and highly specific audio representations to achieve optimal results.

In recent years, models based on language models (LMs), such as AudioLM \cite{borsos_audiolm_2023} and MusicLM \cite{agostinelli_musiclm_2023}, have gained attention for their robust semantic modeling capabilities. However, these models largely depend on text-based prompts for conditioning, which limits their ability to capture subtle audio nuances. As a result, music producers often struggle to articulate detailed audio characteristics using abstract textual descriptions.

To bridge the gap between advanced generative modeling and intuitive user interaction, our prior work introduced the Generative Sample Map (GESAM), which uses a two-dimensional latent space learned through a VAE to condition a Transformer-based model for drum sound generation \cite{limberg2024mapping}. GESAM provided a more intuitive visual interface compared to textual or categorical descriptors. However, the initial design of GESAM was focused solely on percussion sounds and did not consider pitch control. This work extends the GESAM framework by incorporating pitch conditioning with a pitch-timbre disentangled latent space.

Pitch-controlled neural instrument sound synthesis is motivated by the observation that timbre varies with pitch, making simple pitch-shifting algorithms inadequate and necessitating more advanced methods to capture this dynamic behavior. In prior research, such as \cite{engel_neural_2017, engel_gansynth_2018}, audio samples are encoded as high-dimen-\allowbreak sional vectors, e.g., a 512-dimensional representation. While it is possible to explore the latent space by interpolating between representations of different instruments, as demonstrated in \cite{narita_ganstrument_2023, zhang_hyperganstrument_2024}, the high-dimensional nature of the representations limits their interactivity in practical applications. In contrast, the proposed GESAM framework encodes instruments in a two-dimensional (2D) latent space, offering an interpretable representation for human users and significantly enhancing model interactivity for broader application scenarios.

Building on the aforementioned work, this paper introduces the pitch-conditioned Generative Sample Map (pGESAM), which employs a semi-supervised learning strategy for pitch-conditioned neural instrument sound synthesis. Our approach maps instrument sounds to a two-dimensional timbre latent space, where pitch information is removed through a dedicated neighbor loss along with pitch and instrument id classifiers. A conditional Transformer generator then synthesizes high-quality audio based on specified pitch and timbre parameters. Experiments on the NSynth dataset demonstrate that pGESAM achieves accurate pitch control while enabling intuitive navigation and manipulation of the timbre latent space for expressive sound synthesis.

The code base of our pGESAM framework is open source and available as a GitHub repository \footnote{\href{https://github.com/faressc/pgesam}{https://github.com/faressc/pgesam}}. Our contributions include:

\begin{enumerate} 
    \item Extending the GESAM framework to generate \sloppy pitch-conditioned instrument sounds beyond percussive timbres. 
    \item Introducing a novel two-stage semi-supervised training scheme for disentangling pitch and timbre within the latent representation. 
    \item Demonstrating improvements in pitch accuracy and timbral expressiveness through comprehensive experimental evaluation. 
    \item Providing an interactive web application for intuitive exploration and manipulation of the learned latent space, enhancing practical usability for music creators. 
\end{enumerate}

\section{Related Work}

Recent advances in deep learning have substantially propelled audio synthesis and musical sample generation, with a wide range of methods explored to enhance the quality, diversity, and controllability of synthesized audio. This section reviews relevant methodologies and positions our contributions within this evolving landscape.

Autoregressive models, such as WaveNet \cite{oord_wavenet_2016} and MelNet \cite{vasquez_melnet_2019}, have demonstrated impressive capability in generating high-fidelity audio. However, their computational requirements due to sequential generation limit practical usability in interactive applications. Variational Autoencoders (VAEs), on the other hand, provide efficient latent representations enabling rapid sample generation and exploration \cite{caillon_rave_2021, engel_neural_2017, wu_jukedrummer_2022}. While VAEs effectively encode timbral characteristics, controlling discrete musical attributes, particularly pitch, remains challenging without explicitly designed disentanglement mechanisms. Generative Adversarial Networks (GANs), notably introduced in the audio domain by WaveGAN \cite{donahue_adversarial_2018}, have advanced conditional audio synthesis, enhancing sample realism and enabling targeted attribute manipulation. GANSynth \cite{engel_gansynth_2018} exploits adversarial training to improve music sound synthesis quality. Some subsequent works adopted GANs in various generative tasks and conditional generation in specific domains \cite{drysdale_adversarial_2020, nistal_darkgan_2021, gupta_signal_2021, nistal_drumgan_2022, yeh_exploiting_2022}. Recent GAN-based approaches \cite{narita_ganstrument_2023, zhang_hyperganstrument_2024} specifically address pitch-timbre disentanglement, significantly improving the editability and controllability of synthesized sounds. Despite these advances, GAN-based methods often require intricate training regimes and lack intuitive interaction methods for users unfamiliar with technical controls.

Denoising diffusion probabilistic models have also gained pro-minence for their robust generation quality and efficiency, gradually transforming noise signals into structured audio content \cite{chen_wavegrad_2020, huang_fastdiff_2022, kong_diffwave_2020, yang_diffsound_2023}. Although promising, diffusion models typically do not inherently support intuitive real-time interaction or direct control over disentangled musical attributes.

A noteworthy trend in audio generation research emphasizes intuitive user interaction, achieved by either exploring interpolations within pre-trained latent spaces \cite{engel_neural_2017, engel_gansynth_2018, 9670718, narita_ganstrument_2023, zhang_hyperganstrument_2024} or integrating musical domain knowledge directly into generative models \cite{ramires_neural_2019, tomczak_drum_2020, chandna_loopnet_2021, devis_continuous_2023, zhang_controllable_2023-2, wu_music_2024}. However, few existing approaches simultaneously achieve effective pitch conditioning, intuitive user exploration, and precise disentanglement between pitch and timbre. Our work directly addresses these limitations by extending the Generative Sample Map (GESAM) framework \cite{limberg2024mapping, limberg_transformer-based_2025} to explicitly incorporate pitch conditioning and timbre disentanglement within a unified semi-supervised learning approach. Unlike previous methods, our approach effectively integrates a Transformer model with a carefully structured 2D latent representation derived from a pitch-conditioned VAE. This design enables musicians to intuitively and precisely control pitch and timbral nuances, significantly enhancing expressive possibilities in musical sample generation.

\section{Approach}

We approach timbre and pitch disentanglement for the GESAM framework \cite{limberg2024mapping, limberg_transformer-based_2025} by introducing a specialized neighbor loss along with pitch and instrument id classifiers. Following GESAM, we employ the EnCodec model \cite{defossez2022high} to encode audio samples into embeddings prior to training and decode them back into waveforms after generation.

\begin{figure*}[ht]
  \center
    \includegraphics[width=0.8\textwidth]{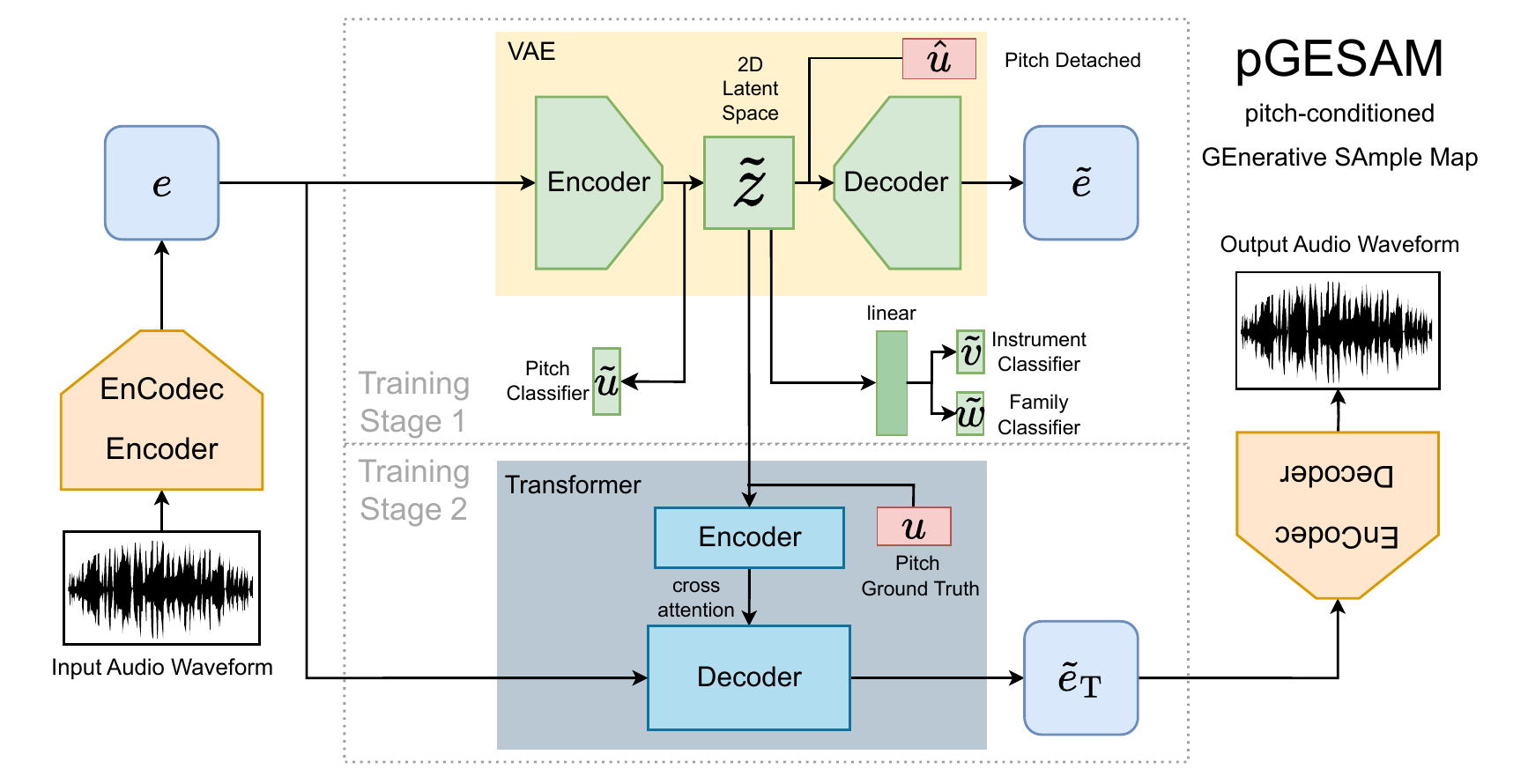}
    \caption{\label{fig:main}{\it Main training paradigm of our approach. }}
\end{figure*}

Our framework operates in two main stages. First, a Variational Autoencoder (VAE) learns a two-dimensional timbre latent space with explicit disentanglement constraints. Second, a Transformer model generates high-quality audio embeddings conditioned on both the learned timbre representations and pitch information. Figure \ref{fig:main} illustrates the complete framework architecture.

\subsection{Stage 1: Variational Autoencoder (VAE) Training}

The goal of the VAE is to create an interpretable 2D representation of timbres, which is then used to condition the transformer in stage 2 and serves as a user interface to explore the landscape of different timbres. The VAE consists of an encoder with five 1D convolutional layers, each with channel sizes $[128, 256, 512, 1024, 2048]$ and a stride of 2, followed by five fully connected layers with sizes $[8192, 4096, 2048, 1024, 512]$. The output of the last layer is fed to two different heads, a regression head and a pitch classification head. The output of the regression head is the mean $\tilde{\mu}$ and logvar $\log(\tilde{\sigma}^2)$ of the two-dimensional timbre latent space. Both are used to sample the latent space vector $\tilde{z}$ using the reparametrization trick. The output of the pitch classification head is a logit vector $\tilde{u}$, which is used to compute the pitch classification loss.

For instrument classification, the latent mean vector $\tilde{\mu}$ is fed into a classification network consisting of fully connected layers with sizes $[2, 4, 8, 16, 32, 64, 32, 64, 128, 256, 512, 1024, N_{\text{inst}}]$, where \sloppy$N_{\text{inst}}$ is the number of instrument ids. Its output is the logit vector $\tilde{v}$. We also use a family classifier to cluster instruments by family. The family classifier's contribution is weighted by an exponentially decaying term that provides stronger influence early in training, gradually diminishing over time. In contrast, the instrument classifier's contribution increases over time (Equation \ref{eq:vae_loss}). The family classifier shares the same input as the instrument classifier and outputs a logit vector, denoted by $\tilde{w}$, with $N_{\text{fam}}$ classes. The family classifier uses dense layers with sizes $[2, 4, 8, 16, 32, N_{\text{fam}}]$.

The VAE's decoder takes as input the concatenation of the predicted timbre latent vectors $\tilde{z}$ and a one-hot encoded pitch class vector $\hat{u}$. The latter is computed by separating the pitch logit vector $\tilde{u}$ from the gradient and then using an argmax operation to obtain the one-hot encoded vector $\hat{u}$. The decoder network consists of five fully connected layers with sizes $[512, 1024, 2048, 4096, 8192]$ and five 1D transposed convolutional layers with channel sizes $[2048, 1024, 512, 256, 128]$ and a stride of 2. The architecture of the VAE is shown in Figure \ref{fig:main}.

To achieve effective disentanglement of pitch and timbre information while ensuring that the latent space is well-structured with macro-clusters of instrument families and micro-clusters of individual instruments, we introduce a seven-component loss function. This loss function balances the objectives of reconstruction quality, regularization, and classification. It is defined as follows:

\begin{align}
 \mathcal{L}_{\text{VAE}} &= \beta_{\text{rec}} \mathcal{L}_{\text{rec}} + \beta_{\text{KL}} \mathcal{L}_{\text{KL}} + \beta_{\text{reg}} \mathcal{L}_{\text{reg}} + \beta_{\text{nei}} \gamma^{\alpha_{\text{nei}}} \mathcal{L}_{\text{nei}} + \beta_{\text{pitch}} \mathcal{L}_{\text{pitch}} \nonumber\\
&+ \beta_{\text{inst}} \gamma^{\alpha_{\text{inst}}} \mathcal{L}_{\text{inst}} + \beta_{\text{fam}} \left(1 - \gamma\right)^{\alpha_{\text{fam}}}\mathcal{L}_{\text{fam}}\label{eq:vae_loss}
\end{align}
where $\mathcal{L}_{\text{rec}}$ is the reconstruction loss, $\mathcal{L}_{\text{KL}}$ is the Kullback-Leibler divergence loss, $\mathcal{L}_{\text{reg}}$ is the regularization loss, $\mathcal{L}_{\text{nei}}$ is the neighbor loss, $\mathcal{L}_{\text{pitch}}$ is the pitch classification loss, $\mathcal{L}_{\text{inst}}$ is the instrument classification loss, and $\mathcal{L}_{\text{fam}}$ is the family classification loss. The scalar hyperparameters $\beta_{\text{rec}}, \beta_{\text{KL}}, \beta_{\text{reg}}, \beta_{\text{nei}}, \beta_{\text{pitch}}, \beta_{\text{inst}}, \beta_{\text{fam}}$ control the relative importance of each loss component.

To promote stable training and proper convergence, we employ a curriculum learning strategy where certain loss components are introduced gradually. The scheduling parameter $\gamma = \frac{i_{\text{epoch}}}{N_{\text{epoch}}}$ represents the training progress, where $i_{\text{epoch}}$ is the current epoch and $N_{\text{epoch}}$ is the total number of training epochs. The neighbor loss and instrument classification loss are gradually weighted up using $\gamma^{\alpha_{\text{nei}}}$ and $\gamma^{\alpha_{\text{inst}}}$ respectively, while the family classification loss is weighted down using $(1-\gamma)^{\alpha_{\text{fam}}}$. The exponents $\alpha_{\text{nei}}, \alpha_{\text{inst}}, \alpha_{\text{fam}}$ control the rate of these scheduling transitions. This approach ensures that the model first learns coarse-grained family distinctions before progressively focusing on fine-grained instrument-level and neighborhood structure. The VAE hyperparameters are summarized in Table \ref{tab:vae_hyperparams}. The training is performed using minibatches of size 256.

\begin{table}[ht]
  \centering
  \caption{Hyperparameters for balancing the VAE loss function}
  \begin{tabular}{|c|c|c|c|c|c|c|c|}
    \hline
    \textbf{Term} & $\mathcal{L}_{\text{rec}}$ & $\mathcal{L}_{\text{KL}}$ & $\mathcal{L}_{\text{reg}}$ & $\mathcal{L}_{\text{nei}}$ & $\mathcal{L}_{\text{pitch}}$ & $\mathcal{L}_{\text{inst}}$ & $\mathcal{L}_{\text{fam}}$ \\
    \hline
    $\beta$ & 0.2 & 0.0039 & 1.0 & 0.6 & 0.07 & 0.12 & 2.0 \\
    $\alpha$ & – & – & – & 0.5 & – & 0.8 & 1.5 \\
    \hline
  \end{tabular}
  \label{tab:vae_hyperparams}
\end{table}

The reconstruction loss employs mean squared error (MSE) between the original and reconstructed audio embeddings:
\begin{align}
 \mathcal{L}_{\text{rec}} = \frac{1}{N} \sum_{i=0}^{N - 1} \left(\tilde{e}_{i} - e_i\right)^2
\end{align}
where $N$ is the minibatch size, $\tilde{e}_{i}$ is the reconstructed audio embedding, and $e_i$ is the original audio embedding for the $i$-th sample.

The Kullback-Leibler divergence loss regularizes the latent space by encouraging distributions to approximate a standard normal distribution $\mathcal{N}(0, I)$:
\begin{align}
 \mathcal{L}_{\text{KL}} = \frac{1}{2} \sum_{i=0}^{N - 1} \left(\tilde{\mu}_i^2 + \tilde{\sigma}_i^2 - \log(\tilde{\sigma}_i^2) - 1\right)
\end{align}
where $\tilde{\mu}_i$ and $\tilde{\sigma}_i$ are the mean and standard deviation of the latent distribution for the $i$-th sample, respectively.

The regularization loss constrains all latent vectors to lie within the unit circle:
\begin{align}
 \mathcal{L}_{\text{reg}} = \frac{1}{N} \sum_{i=0}^{N - 1} \max(0, \|\tilde{\mu}_i\|_2 - 1)
\end{align}
where $\|\tilde{\mu}_i\|_2$ denotes the L2 norm of the latent mean vector.

The neighbor loss promotes structured clustering by applying attractive forces between samples of the same instrument class and repulsive forces between different classes. Inspired by metric learning and the repulsion loss in \cite{repulsion}, it combines:
\begin{align}
\mathcal{L}_{\text{nei}} = \mathcal{L}_{\text{attractive}} + \mathcal{L}_{\text{repulsive}}
\end{align}

The attractive component minimizes distances between latent representations of identical instruments:
\begin{align}
\mathcal{L}_{\text{attractive}} = \frac{\sum_{i=0}^{N-1} \sum_{j=0, j \neq i}^{N-1} S_{ij} \cdot d_{ij}^2}{\sum_{i=0}^{N-1} \sum_{j=0, j \neq i}^{N-1} S_{ij} + \epsilon}
\end{align}

The repulsive component enforces a margin $M$ between different instrument classes:
\begin{align}
\mathcal{L}_{\text{repulsive}} = \frac{\sum_{i=0}^{N-1} \sum_{j=0, j \neq i}^{N-1} (1-S_{ij}) \cdot \max(0, M - d_{ij})^2}{\sum_{i=0}^{N-1} \sum_{j=0, j \neq i}^{N-1} (1-S_{ij}) + \epsilon}
\end{align}
Here, $S_{ij} = \mathbf{1}_{y_i = y_j}$ indicates whether samples $i$ and $j$ share the same instrument class, $d_{ij} = \|\tilde{\mu}_i - \tilde{\mu}_j\|_2$ represents the Euclidean distance between latent mean vectors, $M = 0.25$ is the margin parameter, and $\epsilon = 10^{-6}$ ensures numerical stability. This formulation encourages compact, well-separated instrument clusters in the latent space.

The final three components are the loss functions for the classification heads in the architecture. All three employ cross-entropy loss, calculated as:

\begin{align}
 \mathcal{L}_{\text{class}} = -\frac{1}{N} \sum_{i=0}^{N-1} \sum_{j=0}^{N_{\text{class}}-1} x_{ij} \log(p_{ij})
\end{align}
where $N_{\text{class}}$ is the number of classes, $x_{ij}$ is the ground truth one-hot encoded class vector for the $i$-th sample in the minibatch, and $p_{ij} = \text{softmax}(\tilde{x}_{ij})$, with $\tilde{x}_{ij}$ representing the logits from the corresponding classification head.

For the pitch classifier loss $\mathcal{L}_{\text{pitch}}$, we set $x_{i,j} = u_{i,j}$ (the one-hot encoded ground truth pitch class) and $p_{i,j} = \text{softmax}(\tilde{u}_{i,j})$, where $\tilde{u}_{i,j}$ denotes the logits from the pitch classification head. This classifier predicts the pitch class of audio embeddings based solely on the input embedding $e$ and is independent of the timbre latent vectors $\tilde{z}$, thereby encouraging disentangled pitch and timbre representations.

The instrument and family classifiers follow analogous formulations, using $x_{i,j} = v_{i,j}$ and $x_{i,j} = w_{i,j}$ (the one-hot encoded ground truth instrument and family classes) with corresponding softmax probabilities $p_{i,j} = \text{softmax}(\tilde{v}_{i,j})$ and $p_{i,j} = \text{softmax}(\tilde{w}_{i,j})$, respectively. The instrument classifier ensures that latent vectors from identical instrument ids are clustered by pitch, while the family classifier promotes clustering of instruments within the same family.

\subsection{Stage 2: Pitch/Timbre-Conditioned Transformer}

The second stage employs a Transformer model that generates high-quality audio embeddings from the VAE's learned timbre latent space and pitch conditioning. The Transformer follows an encoder-decoder architecture where the encoder processes latent representations through multi-head self-attention and feed-forward networks, conditioning the decoder via cross-attention. The decoder employs masked self-attention, cross-attention to encoder outputs, feed-forward processing, and a final linear projection to produce audio embeddings.

The encoder receives samples from the predicted timbre latent distributions $\tilde{z} = \tilde{\mu} + \tilde{\sigma} \cdot \epsilon \cdot \beta$ from the VAE, where $\epsilon \sim \mathcal{N}(0, 1)$ and $\beta = 0.01$ controls sampling stochasticity. The timbre latent vector $\tilde{z}$ is linearly projected to match the transformer's dimensionality, while the pitch ground truth $u$ is embedded to the model dimension. These projected representations are positionally encoded, concatenated, and processed by the encoder, whose output provides keys and values for the decoder's cross-attention mechanism.

The decoder receives a sequence of masked audio embeddings prefixed with a \texttt{<BOS>} token (initialized as a zero vector). The masking enforces autoregressive behavior by preventing attention to future time steps. These masked embeddings are linearly projected to the model dimension and combined with positional embeddings to encode sequence order. The decoder processes this input through masked self-attention layers and cross-attention to encoder outputs, followed by feed-forward networks and a linear projection that restores the original audio embedding dimensionality. The model trains autoregressively to predict the next audio embedding given previous embeddings and encoder context, using MSE loss between predicted and ground truth audio embeddings.

During generation, users can select a point in the learned 2D latent space along with a pitch class, and the model generates audio embeddings iteratively token by token starting from the \texttt{<BOS>} token. The hyperparameters for both encoder and decoder are detailed in Table \ref{tab:transformer_hyperparams}.

\begin{table}[ht]
  \centering
  \caption{Hyperparameters for the Transformer model}
  \begin{tabular}{|c|c|}
    \hline
    \textbf{Parameter} & \textbf{Value} \\
    \hline
    Model dimension & 512 \\
    Number of layers encoder & 8 \\
    Number of layers decoder & 12 \\
    Number of attention heads & 8 \\
    Feed-forward dimension & 8192 \\
    \hline
  \end{tabular}
  \label{tab:transformer_hyperparams}
\end{table}

\section{Evaluation}

In this section, we present an evaluation of our proposed pitch-conditioned neural instrument sound synthesis method. Using the NSynth dataset \cite{engel_neural_2017}, we evaluate the performance of our approach along several critical dimensions: reconstruction quality, pitch accuracy, and timbre expressiveness. In the evaluation, we compare our transformer model to the VAE decoder as a baseline. In addition, we perform an ablation study to analyze how the proposed loss function components contribute to the successful design of a latent space that exhibits both perceptual coherence and musical interpretability, enabling intuitive manipulation of timbral characteristics while maintaining pitch accuracy and ensuring seamless interpolation between different instrument voices for expressive performance applications.

\subsection{Dataset}

For our experiments, we use the NSynth dataset \footnote{\href{https://magenta.tensorflow.org/datasets/nsynth}{https://magenta.tensorflow.org/datasets/nsynth}}, which contains a large collection of musical instrument sounds. The dataset contains a variety of instruments from 11 different families, each with multiple samples at different pitches and velocities. For our experiments, we limit the dataset to a subset of pitches ranging from MIDI note number 48 to 72, which corresponds to the range C2 to C4, and we use the samples with a velocity of 100 only.

The dataset is divided into training, validation, and test sets, with 94.52\% of the samples used for training, 4.14\% for validation, and 1.34\% for testing. For each dataset, we limited the number of instrument ids per instrument family to 10, resulting in a maximum of 110 instrument ids. The instrument ids in the datasets do not overlap. All samples are mono audio files with a sample rate of 16kHz and a duration of 4 seconds.

The audio samples are upsampled and then encoded using the EnCodec 24kHz model from MetaAI \footnote{\href{https://huggingface.co/facebook/encodec_24khz}{https://huggingface.co/facebook/encodec\_24khz}}, which compresses the audio samples into a lower-dimensional representation. The EnCodec model is trained on a large dataset of audio samples and is capable of producing high-quality audio embeddings. In our case, the compression ratio of the EnCodec model is 2.5:1 because we skip the vector quantization step, which means that the audio embeddings are 2.5 times smaller than the original audio samples. The audio embeddings are then used as input to the VAE and Transformer models.

\subsection{Reconstruction Quality}

To evaluate the audio quality of the proposed model, we process audio samples from the dataset to compute embeddings, which are then re-synthesized using both the VAE model and the VAE-conditioned Transformer model. As a comparison metric, we use the Mean Squared Error (MSE) between these generated audio embeddings and the corresponding ground truth embeddings. This evaluation is performed on both the training and test sets of the NSynth dataset (cf. Table \ref{tab:recon_error_plain}). To ensure consistency, both the VAE and Transformer models are evaluated on the same set of instruments.


\begin{table}[ht]
  \centering
  \caption{Mean Absolute Error (MAE) of reconstruction}
  \begin{tabular}{|c|c|c|}
    \hline
    \textbf{Dataset} & \textbf{VAE} & \textbf{Transformer} \\
    \hline
    Train & 1.14\text{e-}3 & 4.90\text{e-}5 \\
    Test  & 1.02\text{e-}3 & 1.49\text{e-}3 \\
    \hline
  \end{tabular}
  \label{tab:recon_error_plain}
\end{table}

\begin{figure*}[ht]
\center
\subfloat[Original Embedding]{
  \includegraphics[width=0.3\textwidth]{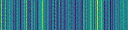}
}
\subfloat[VAE Generated]{
  \includegraphics[width=0.3\textwidth]{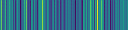}
}
\subfloat[Transformer Generated]{
  \includegraphics[width=0.3\textwidth]{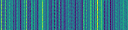}
}
\caption{\label{fig:emb_comparison}{\it Visualization of the latent space with different model configurations. }}
\end{figure*}

The Transformer model achieves higher accuracy on the training set, but exhibits signs of overfitting, as indicated by the higher test loss compared to the VAE. When comparing generated embeddings, it becomes evident that the VAE struggles to reconstruct fine-grained structures: the generated embeddings tend to appear rather flat compared to the ground truth (cf. Figure \ref{fig:emb_comparison}). Listening to the generated audio samples further highlights the importance of these fine structural details for perceived sound quality.

\subsection{Pitch Accuracy}

Generating accurate pitch is essential for neural instrument sound synthesis models. To assess the pitch accuracy of the proposed model, we follow a similar approach as in the previous subsection by reconstructing audio embeddings using both the VAE and Transformer models. The reconstructed audio embeddings are then fed into the VAE encoder, where the trained pitch classifier evaluates the pitch accuracy of the generated samples. We use 0/1 accuracy to compare the generated pitches with the ground truth pitch annotated in the dataset for both the training and test sets (cf. Table \ref{tab:pitch_accuracy}).


\begin{table}[ht]
  \centering
  \caption{Pitch classification 0/1 accuracy}
  \begin{tabular}{|c|c|c|c|}
    \hline
    \textbf{Dataset} & \textbf{GT} & \textbf{VAE} & \textbf{Transformer} \\
    \hline
    Train & 1.00 & 0.112 & 1.00 \\
    Test  & 0.755 & 6.91\text{e-}2 & 0.996 \\
    \hline
  \end{tabular}
  \label{tab:pitch_accuracy}
\end{table}

To ensure the pitch classifier's validity, we first computed the pitch accuracy using the original audio embeddings from the dataset, which yielded a very high accuracy, which confirms the reliability of the pitch classifier in the evaluation. From the experiment results, we observe that the Transformer model is able to generate pitch-accurate samples with perfect accuracy on the training set and over 99\% accuracy on the test set. In contrast, the VAE generates only about 11\% of the samples with the correct pitch. This lower accuracy may result from the VAE's inability to capture fine-grained details, which might be crucial for encoding pitch information effectively in the embedding.

\subsection{Pitch-Timbre Disentanglement} \label{sec:pitch-timbre-disentanglement}

\begin{figure}[ht]
\centerline{\includegraphics[width=0.5\textwidth]{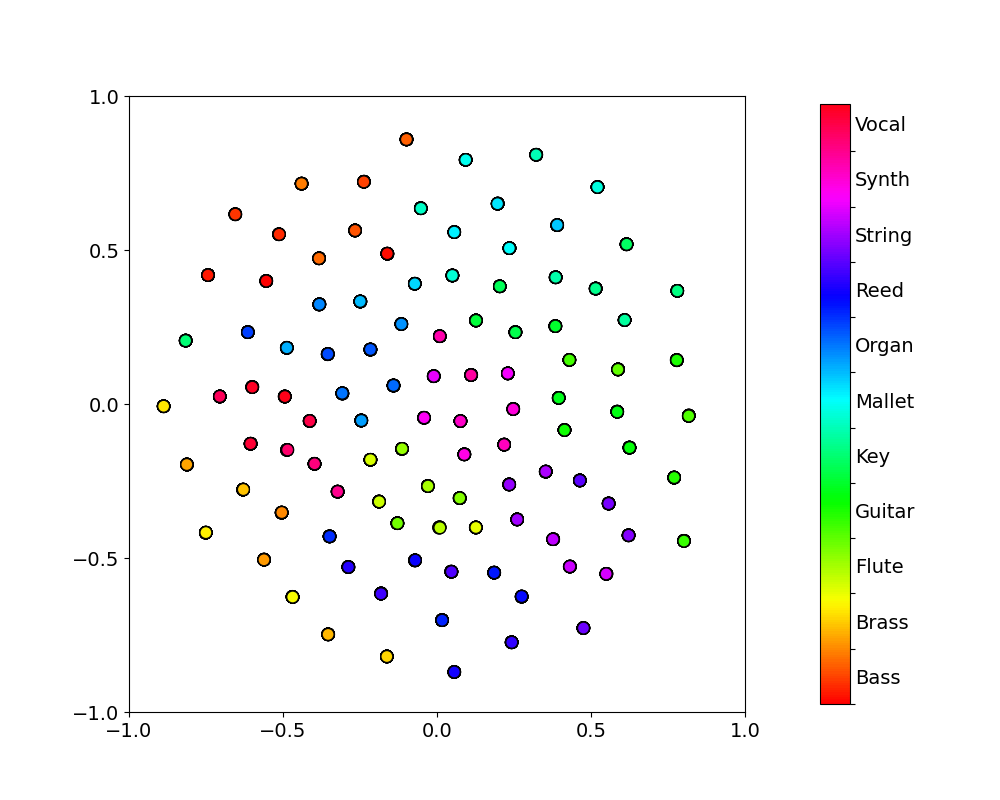}}
\caption{\label{fig:latent_space}{\it Visualization of the latent space of the proposed VAE model. The plot shows the predicted timbre latent mean vectors $\tilde{\mu}$ for all samples in the training set. Each point represents a sample. Instruments of the same family share a base color; the offsets in the color spectrum indicate different instrument ids.}}
\end{figure}

To evaluate the disentanglement of pitch and timbre information in the learned latent space of the VAE, we have first conducted a qualitative analysis of the latent space. We visualize the latent space by plotting the predicted timbre latent mean vectors $\tilde{\mu}$ in a two-dimensional scatter plot, where each point represents a sample in the latent space. The resulting plot is shown in Figure \ref{fig:latent_space}. We can observe very tight clusters of latent mean vectors that are evenly distributed within the unit circle. These clusters represent different instrument ids (encoded in a color transition), with each cluster containing samples of the same instrument at different pitches, demonstrating successful pitch-timbre disentanglement. In the figure, the clusters are extremely tight - indicating a strong pitch-timbre disentanglement - which makes them appear as single points rather than distinct scatter symbols.

In a second quantitative evaluation step for the disentanglement of pitch and timbre information, we examine the variance of the latent mean vectors by instrument id $V_{\text{inst}}$. Since our dataset contains only one velocity level, different samples with the same instrument id represent a different pitch. Assuming that the pitch and timbre information are disentangled and the latent space represents only the timbre information, the variance of the latent mean vectors for samples with the same instrument id should be small. We compute the variance of the latent mean vectors component-wise for each instrument id, and then average them over all instrument ids:
\begin{align}
  V_{\text{inst}} = \frac{1}{N_{\text{inst}}} \sum_{i=0}^{N_{\text{inst}}-1} \frac{1}{N_i} \sum_{j=0}^{N_i-1} (\tilde{\mu}_{ij} - \bar{\tilde{\mu}}_i)^2
\end{align}
Here, $N_{\text{inst}}$ is the number of instrument ids, $N_i$ is the number of samples for the $i$-th instrument id, $\tilde{\mu}_{ij}$ is the latent mean vector for the $j$-th sample of the $i$-th instrument id, and $\bar{\tilde{\mu}}_i$ is the mean of the latent mean vectors for the $i$-th instrument id.

For reference, we compare $V_{\text{inst}}$ to the variance of the predicted latent mean vectors of samples within the same pitch class $V_{\text{pitch}}$. For ideal user control, the latent mean vectors for each pitch class, representing projections of the different instrument ids, should form a uniform distribution on a unit circle. It is calculated as follows:
\begin{align}
  V_{\text{pitch}} = \frac{1}{N_{\text{pitch}}} \sum_{i=0}^{N_{\text{pitch}}-1} \frac{1}{N_i} \sum_{j=0}^{N_i-1} (\tilde{\mu}_{ij} - \bar{\tilde{\mu}}_i)^2 
\end{align}
where $N_{\text{pitch}}$ is the number of pitch classes, $N_i$ is the sample count for pitch class $i$, $\tilde{\mu}_{ij}$ is the latent mean vector for sample $j$ of pitch class $i$, and $\bar{\tilde{\mu}}_i$ is the mean latent mean vector for pitch class $i$.

We compute the variance for both training and test datasets are shown in Table \ref{tab:variance}. The $V_{\text{inst}}$ values are six orders of magnitude smaller than $V_{\text{pitch}}$ in the training dataset, demonstrating effective disentanglement between pitch and timbre information. For the test dataset, while $V_{\text{inst}}$ remains smaller than $V_{\text{pitch}}$, the difference is less pronounced, indicating that the model generalizes reasonably to unseen data while maintaining separation between pitch and timbre characteristics.

The $V_{\text{pitch}}$ values range between 0 and 0.25 for both datasets, aligning with theoretical expectations: while the Kullback-Leibler divergence loss encourages latent mean vectors toward the origin, the regularization and repulsion losses promote uniform distribution of latent vectors within the unit circle, approaching a variance of 0.25 per component.

\begin{table}[ht]
  \centering
  \caption{Variance analysis of latent mean vectors by instrument and pitch for the training and test datasets of the proposed VAE model.}
  \begin{tabular}{|c|c|c|}
    \hline
    \textbf{Dataset} & $[V_{\text{inst, $x$}}, V_{\text{inst, $y$}}]$ & $[V_{\text{pitch, $x$}}, V_{\text{pitch, $y$}}]$ \\
    \hline
    Train & $[1.13\text{e-}7, 1.00\text{e-}7]$ & $[0.179, 0.179]$ \\
    Test & $[2.40\text{e-}2, 2.83\text{e-}2]$ & $[0.136, 0.123]$ \\
    \hline
  \end{tabular}
  \label{tab:variance}
\end{table}

\subsection{Ablation Study}

\begin{figure*}[ht]
\center
\subfloat[No KL loss]{
  \includegraphics[width=0.24\textwidth]{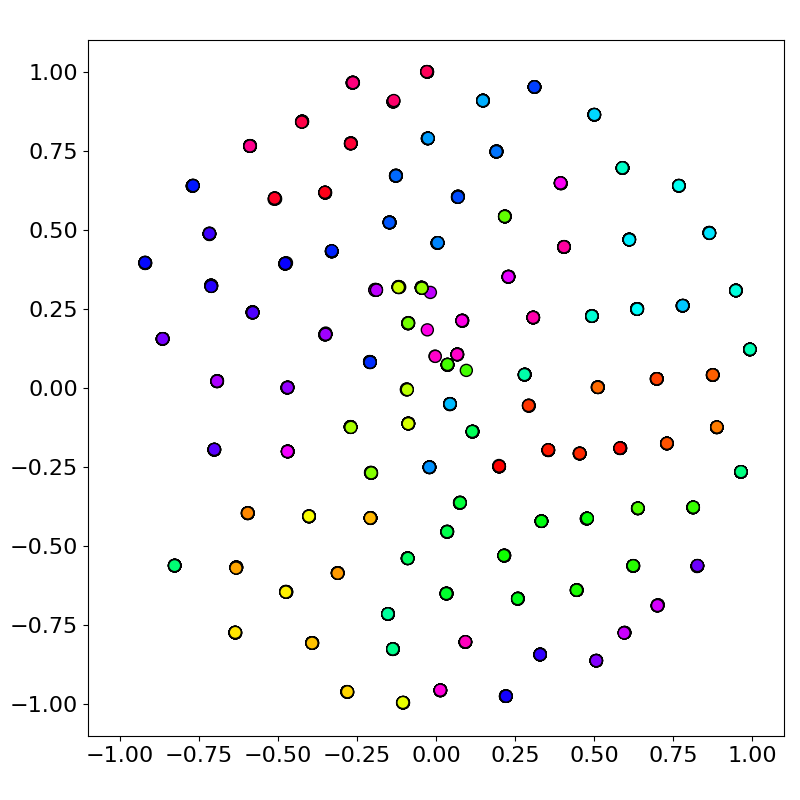}
}
\subfloat[No regularization loss]{
  \includegraphics[width=0.24\textwidth]{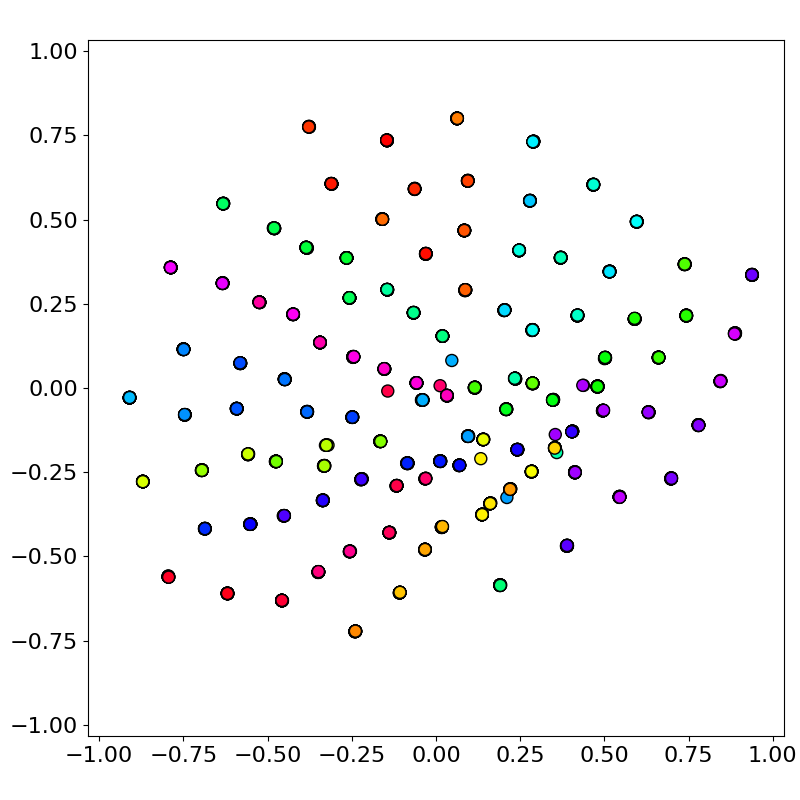}
}
\subfloat[No neighbor loss]{
  \includegraphics[width=0.24\textwidth]{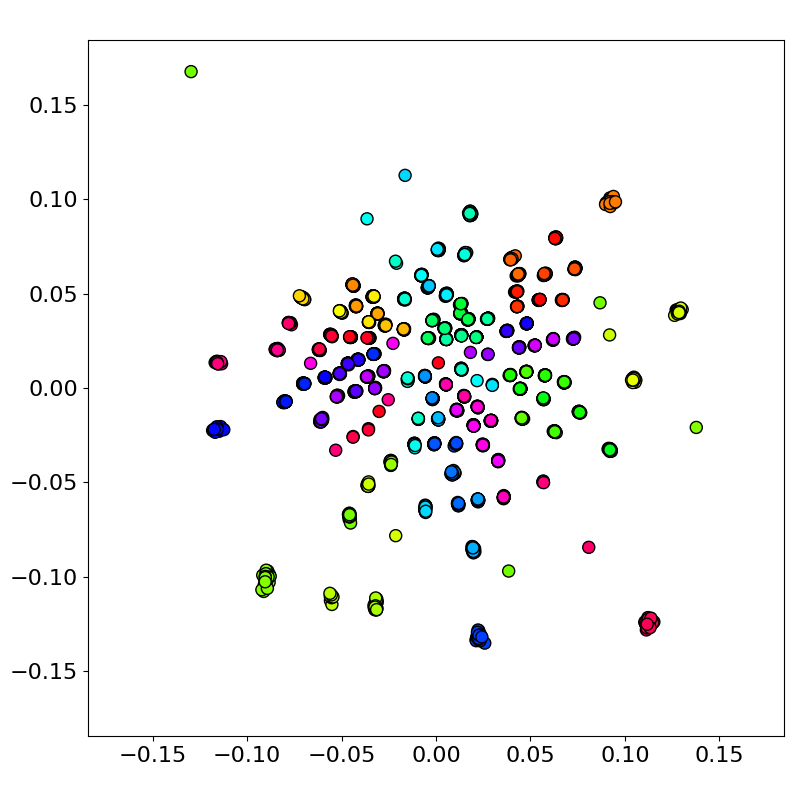}
}
\subfloat[No family classifier]{
  \includegraphics[width=0.24\textwidth]{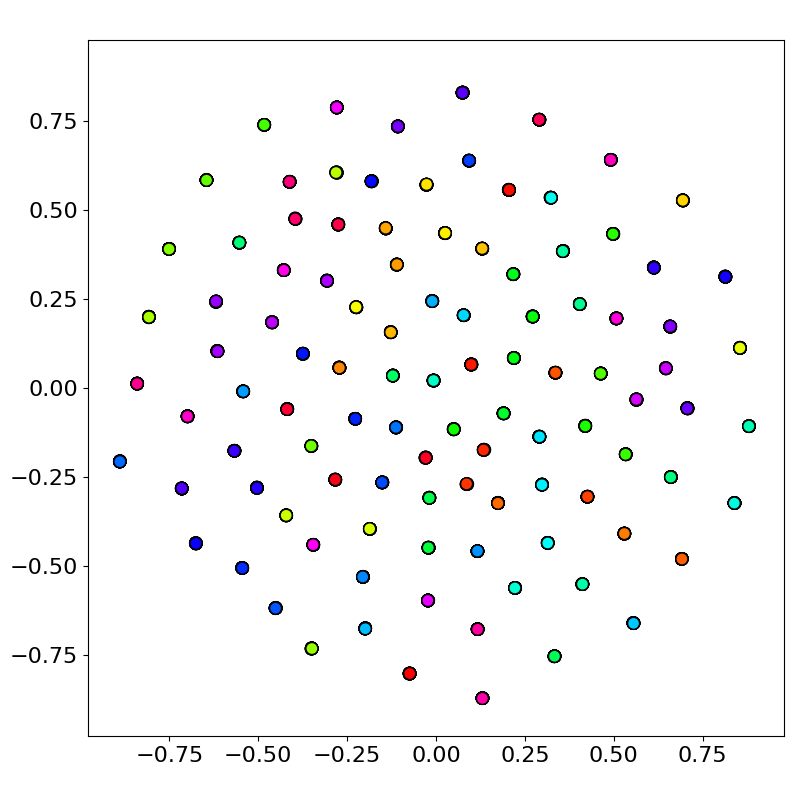}
}
\\
\subfloat{
  \includegraphics[width=0.6\textwidth]{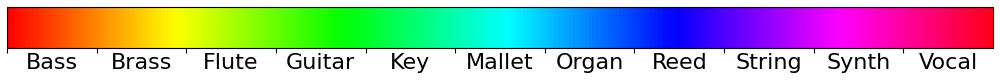}
}
\caption{\label{fig:ablation}{\it Visualization of the latent space with different model configurations. Each subfigure shows the latent space of a VAE trained without one of the proposed components. Each point represents a sample. Instruments of the same family share a base color; the offsets in the color spectrum indicate different instrument ids.}}
\end{figure*}

In this section, we conduct an ablation study to analyze the impact of the proposed loss function components and the family classifier on the formation of the latent space and the disentanglement of pitch and timbre information. We conducted four different experiments to evaluate the effectiveness of the Kullback-Leibler (KL) loss, regularization loss, neighbor loss, and the family classifier. Each experiment involves training a VAE with one of these components removed, allowing us to assess the contributions of each component to the overall performance of the model. Figure \ref{fig:ablation} shows the latent space representation of the trained VAEs. In Table \ref{tab:ablation}, we summarize the results of a variance analysis, similar to Section \ref{sec:pitch-timbre-disentanglement} for each configuration on the training dataset.

\begin{table}[h]
  \centering
  \caption{Variance analysis for each configuration of the ablation study. The variance of the latent mean vectors is shown by instrument and pitch for the training dataset only.}
  \begin{tabular}{|c|c|c|}
    \hline
    \textbf{Model} & $[V_{\text{inst, $x$}}, V_{\text{inst, $y$}}]$ & $[V_{\text{pitch, $x$}}, V_{\text{pitch, $y$}}]$ \\
    \hline
    Baseline & $[1.13\text{e-}7, 1.00\text{e-}7]$ & $[0.179, 0.179]$ \\
    No KL loss & $[2.06\text{e-}5, 8.25\text{e-}7]$ & $[0.229, 0.286]$ \\
    No reg. loss & $[1.93\text{e-}5, 1.49\text{e-}4]$ & $[0.189, 0.125]$ \\
    No nei. loss & $[1.22\text{e-}4, 9.07\text{e-}5]$ & $[2.34\text{e-}3, 2.05\text{e-}3]$ \\
    No family classifier & $[1.58\text{e-}7, 1.76\text{e-}7]$ & $[0.191, 0.173]$ \\
    \hline
    \end{tabular}
    \label{tab:ablation}
\end{table}

The latent space without KL loss (Figure \ref{fig:ablation}a) exhibits more uniformly distributed instrument clusters around the unit circle, as expected since removing KL regularization eliminates the constraint that pulls latent mean vectors toward the origin. This distributional change yields higher $V_{\text{pitch}}$ values compared to the baseline but creates a wider range of relative distances between different instruments, potentially disrupting the semantic organization of the latent space. The increased $V_{\text{inst}}$ values further indicate that pitch-timbre disentanglement degrades without KL regularization. Listening tests confirm these quantitative findings, revealing that samples generated from interpolated points between training data sound less smooth and coherent, demonstrating that KL regularization is essential for maintaining both perceptual continuity and meaningful structural organization in the latent representation.

Without the regularization loss (Figure \ref{fig:ablation}b), the latent space is no longer constrained to the unit circle, resulting in an asymmetric distribution that preferentially spreads along the x-axis. This asymmetry is quantified in the variance analysis, where $V_{\text{pitch, x}} = 0.189$ significantly exceeds $V_{\text{pitch, y}} = 0.125$. While the KL loss prevents completely unbounded growth of the latent representations, the absence of explicit spatial constraints undermines the effectiveness of the neighbor loss, which relies on the limit of the unit circle for meaningful geometric relationships between nearby points in the latent space.

The latent space without the neighbor loss (Figure \ref{fig:ablation}c) shows less uniform spacing between instrument clusters. Due to the KL loss, the latent vectors exhibit a dense population around the origin. While the center becomes densely populated, the other regions of the unit circle remain empty, reducing the effectiveness of latent space exploration. In the variance analysis, we observe that the $V_{\text{pitch}}$ values are significantly lower than those of the baseline model, indicating that the neighbor loss is crucial for encouraging even distribution of instrument timbres across the unit circle.

Finally, we evaluated the latent space representation without the family classifier. Figure \ref{fig:ablation}d reveals the absence of macro-clusters of instrument families, with instruments belonging to the same families now scattered throughout the space rather than grouped together. Interestingly, without the family classifier, the $V_{\text{inst}}$ values exceed those of the baseline model, indicating degraded pitch-timbre disentanglement and suggesting that the family classification task aids in learning semantically meaningful representations.

These ablation experiments collectively demonstrate that each component of our proposed loss function, along with the family classifier, plays a crucial role in forming a well-structured and disentangled latent space that enables intuitive instrument timbre navigation and effective pitch-conditioned synthesis.

\section{Interactivity}
To demonstrate our results, we developed an interactive web demo. Through the demo interface, users can select a point in the 2D latent space to choose a timbre. To select a pitch for generation, users have two options: they can use a slider to select the desired note or use their computer keyboard, which is mapped to a MIDI piano starting with note C3 on the \texttt{a} key. The \texttt{q} key allows users to toggle between two octaves. The interactive demo is available at our website: \href{https://pgesam.faresschulz.com/}{https://pgesam.faresschulz.com/}.

\section{Conclusion}
This paper introduced a pitch-conditioned Generative Sample Map (pGESAM), an interactive framework to synthesize pitch-controlled instrument sounds from an expressive 2D timbre latent space. Our proposed semi-supervised learning strategy effectively disentangles pitch and timbre information, offering users intuitive and creative timbre exploration while preserving the pitch accuracy of the synthesized sounds. We validated our model using the NSynth dataset, demonstrating superior performance in terms of reconstruction quality, pitch accuracy, and timbral expressiveness compared to the baseline method. Through an extended ablation study, we confirmed that the components of our proposed learning objectives, including KL loss, regularization loss, neighbor loss, and family classifier, are crucial for achieving an expressive and interpretable latent space with intuitive interactivity.

To further demonstrate practical usability, we provided an interactive web application, allowing musicians and creators to effortlessly explore and manipulate generated instrument sounds within the intuitive latent space. Future work includes extending the method to more diverse datasets, incorporating additional controllable musical attributes, and realizing variable lengths of synthesized sounds, which will pave the way for further bridging the gap between advanced audio generation models and practical user applications.

\bibliographystyle{IEEEbib}
\bibliography{DAFx25_tmpl} 

\begin{thebibliography}{10}

\bibitem{briot_music_2020}
Jean-Pierre Briot and Fran{\c c}ois Pachet,
\newblock ``Music {{Generation}} by {{Deep Learning}} - {{Challenges}} and {{Directions}},''
\newblock {\em Neural Computing and Applications}, vol. 32, Feb. 2020.

\bibitem{ji_comprehensive_2020-3}
Shulei Ji, Jing Luo, and Xinyu Yang,
\newblock ``A {{Comprehensive Survey}} on {{Deep Music Generation}}: {{Multi-level Representations}}, {{Algorithms}}, {{Evaluations}}, and {{Future Directions}},'' Nov. 2020.

\bibitem{borsos_audiolm_2023}
Zal{\'a}n Borsos, Rapha{\"e}l Marinier, and Damien et~al. Vincent,
\newblock ``{{AudioLM}}: A {{Language Modeling Approach}} to {{Audio Generation}},'' July 2023.

\bibitem{agostinelli_musiclm_2023}
Andrea Agostinelli, Timo~I. Denk, Zal{\'a}n Borsos, Jesse Engel, Mauro Verzetti, Antoine Caillon, Qingqing Huang, Aren Jansen, Adam Roberts, Marco Tagliasacchi, Matt Sharifi, Neil Zeghidour, and Christian Frank,
\newblock ``{{MusicLM}}: {{Generating Music From Text}},'' Jan. 2023.

\bibitem{limberg2024mapping}
Christian Limberg and Zhe Zhang,
\newblock ``Mapping the audio landscape for innovative music sample generation,''
\newblock in {\em ACM International Conference on Multimedia Retrieval}, 2024.

\bibitem{engel_neural_2017}
Jesse Engel, Cinjon Resnick, Adam Roberts, Sander Dieleman, Mohammad Norouzi, Douglas Eck, and Karen Simonyan,
\newblock ``Neural audio synthesis of musical notes with {{WaveNet}} autoencoders,''
\newblock in {\em Proceedings of the 34th {{International Conference}} on {{Machine Learning}} - {{Volume}} 70}, Aug. 2017, pp. 1068--1077.

\bibitem{engel_gansynth_2018}
Jesse Engel, Kumar~Krishna Agrawal, Shuo Chen, Ishaan Gulrajani, Chris Donahue, and Adam Roberts,
\newblock ``{{GANSynth}}: {{Adversarial Neural Audio Synthesis}},''
\newblock in {\em International {{Conference}} on {{Learning Representations}}}, Sept. 2018.

\bibitem{narita_ganstrument_2023}
Gaku Narita, Junichi Shimizu, and Taketo Akama,
\newblock ``{{GANStrument}}: {{Adversarial Instrument Sound Synthesis}} with {{Pitch-Invariant Instance Conditioning}},''
\newblock in {\em {{ICASSP}} 2023 - 2023 {{IEEE International Conference}} on {{Acoustics}}, {{Speech}} and {{Signal Processing}} ({{ICASSP}})}, June 2023, pp. 1--5.

\bibitem{zhang_hyperganstrument_2024}
Zhe Zhang and Taketo Akama,
\newblock ``{{HyperGANStrument}}: {{Instrument Sound Synthesis}} and {{Editing}} with {{Pitch-Invariant Hypernetworks}},'' Jan. 2024.

\bibitem{oord_wavenet_2016}
Aaron van~den Oord, Sander Dieleman, Heiga Zen, Karen Simonyan, Oriol Vinyals, Alex Graves, Nal Kalchbrenner, Andrew Senior, and Koray Kavukcuoglu,
\newblock ``{{WaveNet}}: {{A Generative Model}} for {{Raw Audio}},'' Sept. 2016.

\bibitem{vasquez_melnet_2019}
Sean Vasquez and Mike Lewis,
\newblock ``{{MelNet}}: {{A Generative Model}} for {{Audio}} in the {{Frequency Domain}},'' June 2019.

\bibitem{caillon_rave_2021}
Antoine Caillon and Philippe Esling,
\newblock ``{{RAVE}}: {{A}} variational autoencoder for fast and high-quality neural audio synthesis,'' Dec. 2021.

\bibitem{wu_jukedrummer_2022}
Yueh-Kao Wu, Ching-Yu Chiu, and Yi-Hsuan Yang,
\newblock ``{{JukeDrummer}}: {{Conditional}} beat-aware audio-domain drum accompaniment generation via transformer {{VQ-VAE}},''
\newblock {\em Proceedings of the 23rd ISMIR Conference}, 2022.

\bibitem{donahue_adversarial_2018}
Chris Donahue, Julian McAuley, and M.~Puckette,
\newblock ``Adversarial audio synthesis,''
\newblock {\em Proceedings of ICLR 2019}, 2018.

\bibitem{drysdale_adversarial_2020}
Jake Drysdale, Maciej Tomczak, and Jason Hockman,
\newblock ``Adversarial synthesis of drum sounds,''
\newblock {\em Proceedings of the 23rd International Conference on Digital Audio Effects}, 2020.

\bibitem{nistal_darkgan_2021}
J.~Nistal, S.~Lattner, and G.~Richard,
\newblock ``{{DarkGAN}}: {{Exploiting}} knowledge distillation for comprehensible audio synthesis with {{GANs}},''
\newblock {\em Proceedings of ISMIR Conference}, 2021.

\bibitem{gupta_signal_2021}
Chit. Gupta, Purnima Kamath, and L.~Wyse,
\newblock ``Signal representations for synthesizing audio textures with generative adversarial networks,''
\newblock {\em ArXiv}, vol. abs/2103.07390, 2021.

\bibitem{nistal_drumgan_2022}
J.~Nistal, S.~Lattner, and G.~Richard,
\newblock ``{{DrumGAN}}: {{Synthesis}} of {{Drum Sounds With Timbral Feature Conditioning Using Generative Adversarial Networks}},'' June 2022.

\bibitem{yeh_exploiting_2022}
Yen-Tung Yeh, Bo-Yu Chen, and Yi-Hsuan Yang,
\newblock ``Exploiting pre-trained feature networks for generative adversarial networks in audio-domain loop generation,''
\newblock {\em Proceedings of ISMIR Conference}, 2022.

\bibitem{chen_wavegrad_2020}
Nanxin Chen, Yu~Zhang, Heiga Zen, Ron~J. Weiss, Mohammad Norouzi, and William Chan,
\newblock ``{{WaveGrad}}: {{Estimating Gradients}} for {{Waveform Generation}},'' Oct. 2020.

\bibitem{huang_fastdiff_2022}
Rongjie Huang, Max W.~Y. Lam, and Jun et~al. Wang,
\newblock ``{{FastDiff}}: {{A Fast Conditional Diffusion Model}} for {{High-Quality Speech Synthesis}},'' 2022.

\bibitem{kong_diffwave_2020}
Zhifeng Kong, Wei Ping, Jiaji Huang, Kexin Zhao, and Bryan Catanzaro,
\newblock ``{{DiffWave}}: {{A}} versatile diffusion model for audio synthesis,''
\newblock {\em ArXiv}, vol. abs/2009.09761, 2020.

\bibitem{yang_diffsound_2023}
Dongchao Yang, Jianwei Yu, Helin Wang, Wen Wang, Chao Weng, Yuexian Zou, and Dong Yu,
\newblock ``Diffsound: {{Discrete Diffusion Model}} for {{Text-to-Sound Generation}},''
\newblock {\em IEEE/ACM Transactions on Audio, Speech, and Language Processing}, vol. 31, pp. 1720--1733, 2023.

\bibitem{9670718}
Xuan Shi, Erica Cooper, and Junichi Yamagishi,
\newblock ``Use of speaker recognition approaches for learning and evaluating embedding representations of musical instrument sounds,''
\newblock {\em IEEE/ACM Transactions on Audio, Speech, and Language Processing}, vol. 30, pp. 367--377, 2022.

\bibitem{ramires_neural_2019}
Ant{\'o}nio Ramires, Pritish Chandna, Xavier Favory, Emilia G'omez, and Xavier Serra,
\newblock ``Neural percussive synthesis parameterised by high-level timbral features,''
\newblock {\em ICASSP 2020 - 2020 IEEE International Conference on Acoustics, Speech and Signal Processing (ICASSP)}, 2019.

\bibitem{tomczak_drum_2020}
Maciej Tomczak, Masataka Goto, and Jason Hockman,
\newblock ``Drum synthesis and rhythmic transformation with adversarial autoencoders,''
\newblock {\em Proceedings of the 28th ACM International Conference on Multimedia}, 2020.

\bibitem{chandna_loopnet_2021}
Pritish Chandna, Ant{\'o}nio Ramires, Xavier Serra, and Emilia G'omez,
\newblock ``Loopnet: {{Musical}} loop synthesis conditioned on intuitive musical parameters,''
\newblock {\em ICASSP 2021 - 2021 IEEE International Conference on Acoustics, Speech and Signal Processing (ICASSP)}, pp. 3395--3399, 2021.

\bibitem{devis_continuous_2023}
Ninon Devis, Nils Demerl{\'e}, Sarah Nabi, David Genova, and Philippe Esling,
\newblock ``Continuous {{Descriptor-Based Control}} for {{Deep Audio Synthesis}},''
\newblock in {\em {{ICASSP}} 2023 - 2023 {{IEEE International Conference}} on {{Acoustics}}, {{Speech}} and {{Signal Processing}} ({{ICASSP}})}, June 2023, pp. 1--5.

\bibitem{zhang_controllable_2023-2}
Zhe Zhang, Yi~Yu, and Atsuhiro Takasu,
\newblock ``Controllable lyrics-to-melody generation,''
\newblock {\em Neural Computing and Applications}, vol. 35, no. 27, pp. 19805--19819, Sept. 2023.

\bibitem{wu_music_2024}
Shih-Lun Wu, Chris Donahue, Shinji Watanabe, and Nicholas~J. Bryan,
\newblock ``Music {ControlNet}: {Multiple} {Time}-{Varying} {Controls} for {Music} {Generation},''
\newblock {\em IEEE/ACM Transactions on Audio, Speech, and Language Processing}, 2024.

\bibitem{limberg_transformer-based_2025}
Christian Limberg, Zhe Zhang, and Marc~A. Kastner,
\newblock ``Transformer-{Based} {Audio} {Generation} {Conditioned} by {2D} {Latent} {Maps}: {A} {Demonstration},''
\newblock in {\em {MultiMedia} {Modeling}}, Singapore, 2025, pp. 233--239, Springer Nature.

\bibitem{defossez2022high}
Alexandre D{\'e}fossez, Jade Copet, Gabriel Synnaeve, and Yossi Adi,
\newblock ``High fidelity neural audio compression,''
\newblock {\em arXiv preprint arXiv:2210.13438}, 2022.

\bibitem{repulsion}
Xinlong Wang, Tete Xiao, Yuning Jiang, Shuai Shao, Jian Sun, and Chunhua Shen,
\newblock ``Repulsion loss: Detecting pedestrians in a crowd,''
\newblock {\em CoRR}, vol. abs/1711.07752, 2017.

\end{thebibliography}

\end{document}